# Estimating statistical distributions using an integral identity

Cheng Zhang[†] & Jianpeng Ma[†,‡]

[†]*Applied Physics Program &*

*Department of Bioengineering*

*Rice University*

*Houston, TX 77005*

[‡]*Verna and Marrs McLean Department of Biochemistry and Molecular Biology*

*Baylor College of Medicine*

*One Baylor Plaza, BCM-125*

*Houston, TX 77030*

Abstract

We present an identity for an unbiased estimate of a general statistical distribution. The identity computes the distribution density from dividing a histogram sum over a local window by a correction factor from a mean-force integral, and the mean force can be evaluated as a configuration average. We show that the optimal window size is roughly the inverse of the local mean-force fluctuation. The new identity offers a more robust and precise estimate than a previous one by Adib and Jarzynski [J. Chem. Phys. **122,** 014114, (2005)]. It also allows a straightforward generalization to an arbitrary ensemble and a joint distribution of multiple variables. Particularly we derive a mean-force enhanced version of the weighted histogram analysis method (WHAM). The method can be used to improve distributions computed from molecular simulations. We illustrate the use in computing a potential energy distribution, a volume distribution in a constant-pressure ensemble, a radial distribution function and a joint distribution of amino acid backbone dihedral angles.

# I. Introduction

We present a method for estimating a general statistical distribution from data collected in a molecular simulation. The method is based on an identity and is superior to the common approach of using a normalized histogram, which suffers from either a large noise when the bin size is too small or a systematic bias when the bin size is too large.

Our identity is akin to a previous one derived by Adib and Jarzynski[1] (hence the AJ identity), whereby the distribution density $\rho(x)$ at a point $x$ is estimated from a weighted number of visits to a window surrounding $x$, plus a correction from integrating the derivative of $\rho(x)$. The AJ identity improves over the histogram-based approach not only by eliminating the systematic bias from binning but also by smoothing out the resulting distribution, as the window contains much more data points than a single bin. However, the identity is slightly inconvenient as it neither ensures a positive output, nor determines its optimal parameters.

Here we present a new identity in which we construct a proper correction factor from integrating the "mean force" [the logarithmic derivative of $\rho(x)$], and use it to *divide* the number of visits to a local window to reach an unbiased estimate, as schematically illustrated in Fig. **1**. It offers not only a nonnegative distribution, but also a simple estimate of the optimal window size, as it separates the error contributions from both the histogram and the mean force. The new strategy can also be straightforwardly extended to an arbitrary ensemble and to a joint distribution of multiple variables.

We describe the new identity in section II, present a few numerical examples in section III, and conclude the article in section IV with a few discussions.

## II. Methods

**II.A Integral identity**

We wish to find an expression for the distribution density $\rho(x)$ at $x = x^*$. We first approximate $\rho(x^*)$ by a histogram sum over a local window $(x_-, x_+)$ enclosing $x = x^*$, and then apply a correction factor. Formally, we have

$$\rho(x^*) = \frac{\int_{x_-}^{x_+} \rho(x)\,dx}{\int_{x_-}^{x_+} \rho(x)/\rho(x^*)\,dx},$$

where the numerator $\int_{x_-}^{x_+} \rho(x)\,dx$ counts the fraction of $x$ falling into the window $(x_-, x_+)$, and can thus be measured from the histogram sum over the window.

We then take the logarithm of $\rho(x)/\rho(x^*)$, and write it as an integral of $(\log \rho)'(y)$ as

$$\rho(x)/\rho(x^*) = \exp[\log\rho(x) - \log\rho(x^*)] = \exp\left(\int_{x^*}^{x} (\log\rho)'(y)\,dy\right).$$

So

$$\rho(x^*) = \frac{\int_{x_-}^{x_+} \rho(x)\,dx}{\int_{x_-}^{x_+} \exp\left(\int_{x^*}^{x} (\log\rho)'(y)\,dy\right) dx}. \tag{1}$$

We refer to Eq. (1) as the *fractional identity*. Unlike in the AJ identity[1], the correction here is applied as a divisor instead of additively, see Fig. **1**. Nevertheless, it can be derived as a near-optimal modification of the AJ identity, as shown in Appendix A.

## II.B Mean force from configuration average

The identity Eq. (1) requires a mean force $(\log \rho)'(x)$ in addition to a histogram. Here we shall construct a conjugate force $f_x = f_x(\mathbf{r}^N, \mathbf{s})$, as a function of molecular coordinates $\mathbf{r}^N$ and auxiliary ensemble variables $\mathbf{s}$, such that its ensemble average is the needed $(\log \rho)'(x)$. $f_x$ can be computed from a simulation trajectory every few frames, and the expressions can be found in Eq. (5) and Eq. (4) or (4′).

We first express $x$ as a function $x = X(\mathbf{r}^N, \mathbf{s})$ of both molecular coordinates $\mathbf{r}^N$ and (optionally) some variables $\mathbf{s}$ of the simulation ensemble, e.g., $\mathbf{s}$ can be the volume in an constant-pressure ensemble, or the temperature in a tempering simulation[2,3]. Note, $X(\mathbf{r}^N, \mathbf{s})$ denotes a *function* of $\mathbf{r}^N$ and $\mathbf{s}$, while $x$ its *value*.

The distribution density $\rho(x)$ is defined as

$$\rho(x) = \int \delta(X(\mathbf{r}^N, \mathbf{s}) - x) \, w(\mathbf{r}^N, \mathbf{s}) \, d\mathbf{r}^N d\mathbf{s}, \qquad (2)$$

where $\delta(...)$ is the $\delta$-function, and $w(\mathbf{r}^N, \mathbf{s})$ is the weight for a configuration $\mathbf{r}^N$ and parameters $\mathbf{s}$, e.g., $w(\mathbf{r}^N, \mathbf{s}) \propto \exp[-\beta U(\mathbf{r}^N)]$ in a canonical ensemble [with $U(\mathbf{r}^N)$ being the potential energy and $\beta$ the temperature]. Eq. (2) is properly normalized, for

$$\int \rho(x) dx = \int \left( \int \delta(X(\mathbf{r}^N, \mathbf{s}) - x) dx \right) w(\mathbf{r}^N, \mathbf{s}) \, d\mathbf{r}^N d\mathbf{s} = \int w(\mathbf{r}^N, \mathbf{s}) \, d\mathbf{r}^N d\mathbf{s} = 1.$$

Similarly, the average $\langle A(\mathbf{r}^N, \mathbf{s}) \rangle_x$ at $x$ for any quantity $A(\mathbf{r}^N, \mathbf{s})$ can be defined as

$$\langle A(\mathbf{r}^N, \mathbf{s}) \rangle_x = \frac{\int A(\mathbf{r}^N, \mathbf{s}) \, \delta(X(\mathbf{r}^N, \mathbf{s}) - x) \, w(\mathbf{r}^N, \mathbf{s}) \, d\mathbf{r}^N d\mathbf{s}}{\int \delta(X(\mathbf{r}^N, \mathbf{s}) - x) \, w(\mathbf{r}^N, \mathbf{s}) \, d\mathbf{r}^N d\mathbf{s}}, \qquad (3)$$

where we have again used the $\delta$-function to collect configurations with $X(\mathbf{r}^N, \mathbf{s})$ being $x$, and the denominator is equal to $\rho(x)$. Our objective is to find an quantity $f_x(\mathbf{r}^N, \mathbf{s})$ such that $\langle f_x(\mathbf{r}^N, \mathbf{s}) \rangle_x = (\log \rho)'(x) = \rho'(x)/\rho(x)$.

We now evaluate the derivative of $\rho(x)$ as

$$\rho'(x) = \int -\frac{\partial}{\partial X} \delta(X(\mathbf{r}^N, \mathbf{s}) - x)\, w(\mathbf{r}^N, \mathbf{s})\, d\mathbf{r}^N d\mathbf{s},$$

where we have used $\partial \delta(X - x)/\partial x = -\partial \delta(X - x)/\partial X$.

We proceed by introducing a vector field $\mathbf{v}(\mathbf{r}^N, \mathbf{s})$ such that $\mathbf{v} \cdot \nabla X = 1$, e.g.

$$\mathbf{v}(\mathbf{r}^N, \mathbf{s}) = \frac{\nabla X(\mathbf{r}^N, \mathbf{s})}{\nabla X(\mathbf{r}^N, \mathbf{s}) \cdot \nabla X(\mathbf{r}^N, \mathbf{s})}. \tag{4}$$

More generally, it can be constructed from an arbitrary vector field $\mathbf{Y}$ ($\nabla X \cdot \mathbf{Y} \neq 0$) as

$$\mathbf{v}(\mathbf{r}^N, \mathbf{s}) = \frac{\mathbf{Y}(\mathbf{r}^N, \mathbf{s})}{\nabla X(\mathbf{r}^N, \mathbf{s}) \cdot \mathbf{Y}(\mathbf{r}^N, \mathbf{s})}. \tag{4'}$$

Note $\nabla$ is defined on the joint vector space of both $\mathbf{r}^N$ and $\mathbf{s}$, so $\nabla = (\partial/\partial \mathbf{r}^N, \partial/\partial \mathbf{s})$.

We now insert $1 = \mathbf{v} \cdot \nabla X$ into the integrand and recall that $\delta(X(\mathbf{r}^N, \mathbf{s}) - x)$ depends on $\mathbf{r}^N$ and $\mathbf{s}$ only through $X$. So $\nabla X [\partial \delta(X - x)/\partial X] \to \nabla \delta(X - x)$, and

$$\begin{aligned}\rho'(x) &= \int -\mathbf{v}(\mathbf{r}^N, \mathbf{s}) \cdot \nabla \delta(X(\mathbf{r}^N, \mathbf{s}) - x)\, w(\mathbf{r}^N, \mathbf{s})\, d\mathbf{r}^N d\mathbf{s} \\ &= \int \delta(X(\mathbf{r}^N, \mathbf{s}) - x)\, \nabla \cdot [\mathbf{v}(\mathbf{r}^N, \mathbf{s})\, w(\mathbf{r}^N, \mathbf{s})]\, d\mathbf{r}^N d\mathbf{s} \\ &= \int \delta(X(\mathbf{r}^N, \mathbf{s}) - x)\, w(\mathbf{r}^N, \mathbf{s})\, f_x(\mathbf{r}^N, \mathbf{s})\, d\mathbf{r}^N d\mathbf{s},\end{aligned}$$

where we have integrated by parts to shift the $\nabla$ to the rest of the integrand, and defined a conjugate force $f_x \equiv \nabla \cdot \mathbf{v} + \mathbf{v} \cdot \nabla \log w$. The last step follows from

$$\nabla \cdot (\mathbf{v} w) = \nabla \cdot \mathbf{v}\, w + \mathbf{v} \cdot \nabla w = (\nabla \cdot \mathbf{v} + \mathbf{v} \cdot \nabla \log w)\, w = f_x w.$$

Comparing with Eq. (3), we see that the mean force $(\log \rho)'(x) = \rho'(x)/\rho(x)$ is equal to the average of $f_x(\mathbf{r}^N, \mathbf{s})$ under a fixed $x$,

$$(\log \rho)'(x) = \langle f_x(\mathbf{r}^N, \mathbf{s}) \rangle_x = \langle \nabla \cdot \mathbf{v}(\mathbf{r}^N, \mathbf{s}) + \mathbf{v}(\mathbf{r}^N, \mathbf{s}) \cdot \nabla \log w(\mathbf{r}^N, \mathbf{s}) \rangle_x. \qquad (5)$$

For a canonical ensemble, the second term $\mathbf{v} \cdot \nabla \log w$ is reduced to $-\beta \mathbf{v} \cdot \nabla U$, i.e., the projection of the molecular force $-\nabla U = \mathbf{F}$ to the gradient of $X$ [assuming Eq. (4)], which fits the name "mean force" of $\langle f_x \rangle$.

The above derivation is analogous to that of the dynamic temperature by Rugh[4]. In fact the latter can be derived as a special case of Eq. (5). Consider the canonical ensemble at infinite temperature $\beta \to 0$. The distribution of the total energy $E = H(\mathbf{r}^N)$ ($\mathbf{r}^N$ now denotes a point in the phase space of coordinates and momenta) is proportional to the density of states $g(E)$. The mean force $\partial \log g(E)/\partial E$ should be the intrinsic temperature $\beta(E)$. According to Eq. (5), it equals to $\langle \nabla \cdot \mathbf{v} \rangle_E$, with $\mathbf{v} = \nabla H/(\nabla H \cdot \nabla H)$, we thus recover Rugh's dynamic temperature.

**II.C Optimal window size**

We now determine the two window boundaries $x_-$ and $x_+$ in Eq. (1) such that they minimize the statistical error in $\rho(x^*)$.

We first note that the histogram and mean-force data contribute independently to the numerator and denominator, respectively. How much the two contribute is however controlled by the window size. The output is dominated by the histogram contribution (numerator) with a narrow window, but by the mean-force contribution (denominator) with a wide window. For a narrow window, the denominator is reduced to the window width, and thus the output to a histogram average. At the other extreme, if the window covers the entire domain of $x$, the numerator becomes a constant, and the distribution density is determined entirely by the mean-force integral on the denominator, i.e.,

$\rho(x^*) = \exp\left[\int^{x^*}(\log \rho)'(x)\,dx\right]$ (the lower bound of the integral is to be determined by the normalization).

As the window size increases, the relative error of the numerator decreases as more data points reduces uncertainty, but that of the denominator increases as the error in the mean-force integral accumulates. The sum reaches a minimum at the optimal size.

Quantitatively, the relative error of the numerator $\varepsilon(N)/N$ is $1/\sqrt{N}$, where $N = N(x_-, x_+)$ is the number of independent data points included in the window.

The relative error of the denominator $D$ is harder to compute exactly and thus is estimated from an upper bound. First, since $D$ is an integral of $\exp(\Delta \log \rho)$, the relative error of $D$ is no larger than the maximal relative error of $\exp(\Delta \log \rho)$, or equivalently the maximal absolute error of $\Delta \log \rho$. Next, since $\Delta \log \rho$ itself is an integral of the mean force, i.e., $\Delta \log \rho(x) = \int_{x^*}^{x} \langle f_x \rangle(x')\,dx'$, the maximum is likely to occur at either window boundary $x = x_\pm$. In the discrete version, the integral becomes a sum over bins as $\Delta \log \rho = \sum_i \langle f_x \rangle(x_i)\,\delta x$ with $\delta x$ being the bin size. Its error $[\varepsilon(\Delta \log \rho)]^2$, assuming no correlation among mean force at different bins, is $\sum_i \sigma_f^2(x_i)\,\delta x^2 / \delta n(x_i)$, where $\sigma_f^2(x_i)$ is the variance of the conjugate force at $x_i$, and $\delta n(x_i)$ is the number of independent data points in the bin at $x_i$.

On reaching the optimal window, including one more bin from either edge would keep the combined error $(\varepsilon(N)/N)^2 + (\varepsilon(D)/D)^2$ constant. So

$$-\frac{\delta n(x_\pm)}{N(x_-,x_+)^2} + \frac{\sigma_f^2(x_\pm)}{\delta n(x_\pm)}\delta x^2 = 0,$$

where the first term is the decrease of error due to the increased sample size, while the second is the increase due to the mean-force integration. Thus

$$\frac{\delta n(x_\pm)}{N(x_-,x_+)} = \sigma_f(x_\pm)\,\delta x.$$

For a relatively narrow window, $N \approx \delta n\, w/\delta x$, with $w$ being the window width. If we further replacing $\sigma_f(x_\pm)$ by the local mean $\bar\sigma_f$, then

$$w = \gamma/\bar\sigma_f, \tag{6}$$

where $\gamma$ is a heuristic factor which should ideally be 1.0. However, as we overestimate the error of the denominator, Eq. (6) somehow underestimates the optimal window size. In practice, we found the optimal $\gamma$ was somewhere between 1 and 2.

On using Eq. (6), we stress that $\sigma_f(x)$ is the mean-force fluctuation at a *fixed x*, i.e., $\sigma_f^2(x) = \langle f^2\rangle_x - \langle f\rangle_x^2$, and thus is evaluated from data points in the bin containing $x$. However, after the intra-bin calculation, we can then average $\sigma_f(x)$ over a local or global window to improve the precision of $\bar\sigma_f$.

If the mean-force fluctuation is very small, Eq. (6) suggests abandoning the histogram data and switching to a mean-force integration. On the other hand, if the mean force has a very large variance, we should stick to the histogram. Thus the method is effective only if the mean-force fluctuation is small.

**II.D Summary and practical notes**

To summarize the method, we first compute a mean force profile $(\log \rho)'(x)$ at any $x$ from Eqs. (4) [or (4′)] and (5) from a simulation trajectory, then plug it into Eq. (1) to compute the distribution density $\rho(x^*)$. The combined formula looks like

$$\rho(x^*) = \frac{\int_{x_-}^{x_+} \rho(x) dx}{\int_{x_-}^{x_+} \exp\left[\int_{x^*}^{x} \left\langle \nabla \cdot \left(\frac{\nabla X}{\nabla X \cdot \nabla X}\right) + \frac{\nabla X \cdot \nabla \log w}{\nabla X \cdot \nabla X} \right\rangle_x dy\right] dx}, \quad (7)$$

and the window $(x_-, x_+)$ size is determined by Eq. (6).

Practically, we shall assume both histogram and mean force data are collected by bins of size $\delta x$. We evaluate the double integral $\int_{x_-}^{x_+} \exp\left(\int_{x^*}^{x} (\log \rho)'(y) dy\right) dx$ in Eq. (1) by first computing $v(x, x^*) = \int_{x^*}^{x} (\log \rho)'(y) dy$, then $\int_{x_-}^{x_+} \exp[v(x, x^*)] dx$. The step size for numerical integration is always the bin size $\delta x$. For the inner integral, $(\log \rho)'$ is computed as the bin-averaged value, and $v(x, x^*)$ at every bin boundary $x$ as $v(x, x^*) \approx \sum_{x_i = x^*}^{x} (\log \rho)'(x_i) \delta x$. The outer integral is then evaluated by the trapezoidal rule as

$$\int_{x_-}^{x_+} f(x) dx \approx \left[\frac{f(x_-)}{2} + f(x_- + \delta x) + \cdots + f(x_+ - \delta x) + \frac{f(x_+)}{2}\right] \delta x,$$

where $f(x) = \exp[v(x, x^*)]$.

Since we usually need $\rho(x^*)$ at many $x^*$, we pre-compute $(\log \rho)'(y)$ and then $v(x, x_O) = \int_{x_O}^{x} (\log \rho)'(y) dy$ for some $x_O$. In this way, $v(x, x^*) = v(x, x_O) - v(x^*, x_O)$ can be quickly retrieved in evaluating $\rho(x^*)$ at any $x^*$.

Although the statistical error of $(\log \rho)'(y)$ depends on the bin size, that of the integral $v(x, x^*)$, and hence $\rho(x^*)$, does not. To see this, consider a small bin of width $\delta x$ and $\delta n$ data points. If the standard deviation of $(\log \rho)'(y)$ is $\sigma$, then the error of the bin is $\sigma^2/\delta n$. Its contribution to the error of $v(x, x^*)$ should be multiplied by $\delta x^2$ and is thus $\sigma^2 \delta x^2/\delta n$. If we now split the bin into two, then each sub-bin has roughly $\delta n/2$ points, each with an error of $2\sigma^2/\delta n$. But the contribution to the error of $v(x, x^*)$ from the two new bins, $(2\sigma^2/\delta n)(\delta x/2)^2 \times 2 = \sigma^2 \delta x^2/\delta n$, remains the same.

Practically, Eq. (5) may fail if a bin is empty. In this case, we symmetrically enlarge the bin to a window such that it contains at least one data point, then use Eq. (5).

**II.E Extension to weighted histogram analysis method**

We now extend Eq. (1) to a composite distribution, i.e. a superposition of several distributions under different conditions, which can result from independent simulations or an extended-ensemble simulation, such as a tempering simulation (either simulated[2] or parallel[3] tempering). For concreteness, we assume that the individual distributions are canonical ones $\rho(x, \beta_i)$ at different temperatures $\beta_i$ (with $i$ being its label).

The aim is to estimate the distribution $\rho(x, \beta)$ at some $\beta$, which needs not to be one of the $\beta_i$'s. We show that the standard routine, weighted histogram analysis method (WHAM)[5] can be improved by using the mean-force information.

We first generalize Eq. (1) to

$$\rho(x^*, \beta) = \frac{\sum_i N_i \int_{x_-}^{x_+} \rho(x, \beta_i) dx}{\sum_i N_i \int_{x_-}^{x_+} \frac{\rho(x, \beta_i)}{\rho(x^*, \beta)} dx},$$

where $N_i$ is the total number of independent data points from the simulation at the temperature $\beta_i$, and the sum is carried over different temperatures $\beta_i$. To proceed, we simultaneously multiple and divide $\rho(x^*, \beta_i)$ in the denominator,

$$\int_{x_-}^{x_+} \frac{\rho(x, \beta_i)}{\rho(x^*, \beta)} dx = \frac{\rho(x^*, \beta_i)}{\rho(x^*, \beta)} \int_{x_-}^{x_+} \frac{\rho(x, \beta_i)}{\rho(x^*, \beta_i)} dx,$$

where the $x$-independent $\rho(x^*, \beta_i)/\rho(x^*, \beta)$ has been moved out of the integral. We then convert $\rho(x, \beta_i)/\rho(x^*, \beta_i)$ to the mean force integral at a fixed $\beta_i$ as before,

$$\rho(x^*, \beta) = \frac{\sum_i N_i \int_{x_-}^{x_+} \rho(x, \beta_i) dx}{\sum_i \left( N_i \frac{\rho(x^*, \beta_i)}{\rho(x^*, \beta)} \int_{x_-}^{x_+} \exp[\int_{x^*}^{x} (\log \rho)'(y, \beta_i) dy] dx \right)}. \quad (8)$$

For example, in a canonical ensemble, $w(U, \beta) = \exp(-\beta U)/Z(\beta)$ with $Z(\beta)$ being the partition function. So the potential energy $U$ distribution

$$\rho(U^*, \beta) = \frac{\sum_i N_i \int_{U_-}^{U_+} \rho(U, \beta_i) dU}{\sum_i \left( N_i \frac{\exp(-\beta_i U^*)/Z(\beta_i)}{\exp(-\beta U^*)/Z(\beta)} \int_{U_-}^{U_+} \exp\left[ \int_{U^*}^{U} \left( \langle \nabla \cdot \mathbf{v} \rangle_{U'} - \beta_i \right) dU' \right] dU \right)}, \quad (9)$$

where $\mathbf{v} = \nabla U/(\nabla U \cdot \nabla U)$. The regular WHAM[5] is recovered with an infinitesimal window $U_- = U_+ = U^*$. Generally, Eq. (8) improves the histogram method by using the mean force data, e.g., the dynamic temperature $\langle \nabla \cdot \mathbf{v} \rangle_{U'}$ here. Note since $\langle \nabla \cdot \mathbf{v} \rangle_{U'}$ is the same under any temperature $\beta_i$, its data from different temperatures can be combined.

## III. Examples

### III.A Potential energy distribution

We first compute a potential energy distribution $\rho(U)$ in a canonical ensemble, in which $w(\mathbf{r}^N, \beta) = \exp[-\beta U(\mathbf{r}^N)]/Z(\beta)$ with $\beta$ being the reciprocal temperature, and $Z(\beta)$ being the partition function. Eqs. (1) and (5) become

$$\rho(U^*) = \frac{\int_{U_-}^{U_+} \rho(U) dU}{\int_{U_-}^{U_+} \exp\left(\int_{U^*}^{U} (\log \rho)'(y) dy\right) dU},$$

$$(\log \rho)'(U) = \langle \nabla \cdot \mathbf{v} - \beta \rangle_U = \left\langle \nabla \cdot \left(\frac{\nabla U(\mathbf{r}^N)}{\nabla U(\mathbf{r}^N) \cdot \nabla U(\mathbf{r}^N)}\right)\right\rangle_U - \beta.$$

Since $(\log \rho)'(U)$ is the difference between the dynamic temperature[4,6] $\langle \nabla \cdot \mathbf{v} \rangle$ and the simulation temperature $\beta$, the distribution peaks at $(\log \rho)'(U) = 0$ where the two cancel.

We performed a molecular dynamics (MD) simulation on a 256-particle Lennard-Jones system under a smoothly switched potential (see Appendix C, $r_s = 2.0$ and $r_c = 3.0$). The temperature $\beta = 1.0$, density $\rho = 0.8$, time step $\Delta t = 0.002$. Velocity rescaling was used as the thermostat[7] with a time step 0.01.

From a single trajectory of $10^7$ steps, we constructed two samples. In the test sample, $10^4$ frames were collected every 1000 steps; in the reference one, all frames were used. The setting avoided possible sampling inaccuracy to affect the comparison of different methods, as explained in Appendix D. The bin size for histogram-like data was always $\delta U = 0.1$. Unless specified otherwise, the test sample was used.

We first demonstrate the use of the fractional identity Eq. (1) with a fixed symmetric window of size $\Delta U = U_+ - U_- = 12.4$, a value determined from Eq. (6)

($\gamma = 1.0$). The mean force $(\log \rho)'(U)$ was computed from a single bin at $U$. As shown in Fig. **2**(a), the resulting distribution was much smoother than the histogram (which was calculated from the number of visits to each bin). For comparison, we computed the result from the AJ identity with the same window size. Though the results were generally similar, the AJ identity sometimes yielded negative values at the two edges, while the fractional identity appeared to be more robust and closer to the reference.

To show the gain from the integral identity approach, we define a KS difference as $\Delta_{KS} = (\sqrt{N} + 0.11 + 0.12/\sqrt{N})\Delta_{CDF}$ [which is commonly used in the Kolmogorov-Smirnov (KS) test for detecting the difference between two distributions[8]], where $N$ is the sample size, and $\Delta_{CDF}$ is the maximal difference between the cumulative distribution function (CDF) $F(x) = \int_{-\infty}^{x} \rho(y)\,dy$ of the resulting distribution and that from the reference. The smaller the quantity is, the more accurate the test distribution is. As the measure is independent of the bin size, it mainly detects the systematic bias in the test distribution instead of the smoothness of distribution density. As the identity is not optimized for the CDF but for the distribution density, the KS difference serves as a stringent test.

The KS differences, computed for the histogram, the fractional identity Eq. (1) and the AJ identity Eq. (A1) are shown in Fig. **2**(b). It is clear that that both identities yielded more accurate distributions than the histogram. We also show there was an optimal window size that minimized the error. However, for the fractional identity, the optimal window size $\Delta U \approx 20.0$ was greater than the value 12.4 given by Eq. (6). Thus, a factor $\gamma \approx 1.5$ was used in other examples. Recall the KS difference scales with $\sqrt{N}$, thus from Fig. **2**(b) we estimated about 20-fold increase of efficiency from using the optimal window. We also notice that with a smaller window, the fractional identity gave better estimates than the AJ identity. This was expected as that the fractional identity is the optimal modification of the AJ identity in this case, as shown in Appendix A. With a larger window, the errors from both identities grew rapidly due to the larger involvement of the mean force data. As the factional identity quickly switched to a pure mean-force-based integral with a large window, its growth was faster. The comparison shows that choosing the window size is crucial to the success of the integral identity, and an overly large window can be counterproductive.

We performed a similar comparison in terms of the entropic distance defined as $D(\rho \| \rho^{\text{ref.}}) = \int_{-\infty}^{\infty} \log[\rho(y)/\rho^{\text{ref.}}(y)] \rho(y) dy$ for $\rho(x)$ and the reference $\rho^{\text{ref.}}(x)$. For the AJ identity, in case $\rho(x) < 0$, zero was assumed. Unlike the KS difference, this quantity directly compares the distribution densities. As shown in Fig. **2**(c), the fractional identity consistently produced a small entropic distance than the AJ identity, suggesting an improved smoothness. Similar to the previous case (KS distance), the error from the entropic distance also had a minimal at $\Delta U \approx 20.0$.

We now demonstrate the mean-force-improved weighted histogram analysis method (WHAM) introduced in Sec. II.E. In this case, we performed additional simulations at two neighboring temperatures $T = 0.8$ and $T = 1.2$. The reweighted distributions to $T = 1.0$ from both the original WHAM and the improved version are shown in Fig. **2**(d), and as expected, the latter was much smoother than the former.

We stress that the identity approach is ensemble-dependent because the mean force depends on the ensemble weight $w$. To show this, we simulated the same system using a regular molecular dynamics without a canonical thermostat, i.e., we targeted a microcanonical ensemble with a constant total energy. In the ensemble, the weight for a configuration, after averaging out momentum components, is

$w(\mathbf{r}^N) \propto \sqrt{K}^{N_f-2} = \sqrt{E_{tot} - U(\mathbf{r}^N)}^{N_f-2}$, where $N_f$ is the number of degrees of freedom, and $K$, $U$ and $E_{tot}$ are the kinetic, potential and total energy, respectively. The mean force is accordingly $(\log \rho)'(U) = \langle \nabla \cdot \mathbf{v} \rangle_U - \frac{\frac{1}{2}N_f - 1}{E_{tot} - U}$, where $N_f = 3N - 6$, and the constant reference temperature $\beta$ in the canonical ensemble is changed to an energy-dependent term $(\frac{1}{2}N_f - 1)/(E_{tot} - U)$.

The microcanonical-ensemble simulation was similar to the canonical-ensemble one. During equilibration, the kinetic energy was scaled regularly to match $\bar{T} \approx 1.0$, and was kept as a constant afterwards ($E_{tot} = -932$). As shown in Fig. **2**(d), the distributions and mean forces (lower inset) from the two ensembles differed considerably, whereas the dynamic temperature $\langle \nabla \cdot \mathbf{v} \rangle$ (upper inset) matched. This example shows the importance of applying the correct formula for the mean force.

**III.B Volume distribution**

In the second example, we compute a volume distribution $\rho(V)$ in an isothermal-isobaric (i.e., constant temperature and pressure) ensemble[9,10]. Unlike the previous case, the volume $V$ is not a function of system coordinates, but an additional variable in the ensemble weight $w(\mathbf{r}^N, V)$. Particularly, it serves as a scaling factor that translates the reduced (0 to 1) coordinates $\mathbf{R}_i$ to the actual ones as $\mathbf{r}_i = \sqrt[3]{V}\mathbf{R}_i$. In terms of reduced coordinates, the ensemble weight can be written as

$$w(\{\mathbf{r}_i\},V)\, d\mathbf{r}^N dV \propto \exp[-\beta U(\{\sqrt[3]{V}\mathbf{R}_i\}) - \beta p V]\, V^N d\mathbf{R}^N dV,$$

where $\beta$ and $p$ are the reciprocal temperature and the pressure, respectively.

According to Eq. (5), the conjugate force $f_V$ is reduced to $\partial \log w/\partial V$ in this case (the vector field $\mathbf{v}$ is the unit vector along the direction of the parameter $V$, so $\nabla \cdot \mathbf{v} = 0$, and $\mathbf{v}\cdot\nabla = \partial/\partial V$). Thus,

$$\rho(V^*) = \frac{\int_{V_-}^{V_+} \rho(V)dV}{\int_{V_-}^{V_+} \exp\left(\int_{V^*}^{V}(\log\rho)'(y)dy\right)dV},$$

$$(\log\rho)'(V) = \frac{N}{V} - \frac{\langle \beta\mathbf{r}\cdot\nabla U\rangle_V}{3V} - \beta p = \beta\left[\langle p_c(\mathbf{r}^N)\rangle_V - p\right]$$

The latter represents the difference between the averaged virial pressure $p_c(\mathbf{r}^N) \equiv (N + \beta\langle \mathbf{r}\cdot\mathbf{F}\rangle_V/3)/V$ and the simulation pressure $p$.

There is however a subtle distinction between the apparent volume distribution $\rho(V)$ defined above and the actual physical one $\hat{\rho}(V)$, especially if the system is not spatially periodic. It arises from that the partition function $Z(\beta,V)$ in the canonical ensemble counts configurations with the volume *no larger than*, instead of equal to, the volume $V$ of the simulation box. In other words, unless there are particle lying precisely at all six boundary faces of the simulation box, it is always possible to shrink the box slightly without leaving out any particles. In this sense the physical volume of a configuration is usually less than the volume of the box. Thus it is helpful to use the *differential* partition function $Z'$ [10] for all configurations with the volume precisely falling in $(V-dV, V)$ as $Z' \propto (\partial Z/\partial V) dV = \langle p_c \rangle_V Z \, dV$, and accordingly the actual physical volume distribution $\hat{\rho}(V)$ differs from $\rho(V)$ by a factor, i.e., $\hat{\rho}(V) \propto \langle p_c \rangle_V \rho(V)$. The reader is referred to Koper and Reiss[10] for a more thorough discussion.

However, the above correct does not strictly apply to a periodic system, as the test case here. We calculated $\hat{\rho}(V)$ below only to show the computing process by pretending the system were not periodic. A direct sampling according to $\hat{\rho}(V)$ is nonetheless inconvenient, for we have to know $\langle p_c \rangle_V$ in advance. Thus we shall use $\rho(V)$ to populate configurations during simulation, but adjust $\hat{\rho}(V)$ after simulation.

We performed an MD simulation on the 256 Lennard-Jones system using the switched potential (with $r_s = 2.5$ and $r_c = 3.5$). The temperature $T = 1.24$ and pressure $p = 0.115$ is around the critical point. Velocity rescaling was used as the thermostat[7] with a time step 0.01. For the pressure control, Monte Carlo volume moves were tried every two MD steps with a maximal magnitude of $\pm 2.0\%$ of the side length of the box. The trajectory contained $10^7$ steps with the time step $dt = 0.002$.

From the trajectory, we constructed two samples. In the test sample, we selected every one out of every 100 frames. In the reference one, we used all frames. The setting avoided possible sampling inaccuracy to affect the comparison of different methods, as explained in Appendix D. In both case, histogram-like data were collected using a bin width $\delta V = 1.0$. Unless specified otherwise, the test sample was used.

Since the volume changed almost by an order of magnitude, and the mean force fluctuation $\sigma_f \propto 1/V$, a fixed window size was not suitable. Thus we applied Eq. (6) with $\sigma_f$ estimated from a local window, and $\gamma = 1.5$ (heuristic value).

To apply the correction, we further computed $\langle p_c \rangle_V$ from a second integral identity, Eqs. (B1) and (B2), in Appendix B. The window size of the second identity was similarly determined from the local $\sigma_f$ but with $\gamma = 3.0$.

As shown in Fig. **3**, the volume distribution $\rho(V)$ or $\hat{\rho}(V)$ from the fractional identity was smoother than the histogram but still had some roughness. From the inset, we observe that the window size grew linearly with the volume $V$. This example also clearly illustrates the danger of using an overly large window. We show in Fig. **3** the distribution from a pure mean-force integration $\rho(V) = \exp\left[\int^V (\log \rho)'(V')\,dV'\right]$ (which is the limiting case of using an infinite window) manifested a much larger deviation from the reference. The deviation was however not systematic and diminished with the sample size: when calculated from the larger reference sample, the deviation was hardly noticeable, see Fig. **7**(b). The comparison shows choosing a proper window is crucial to the success of the method.

**III.C Radial distribution function**

In the third example, we compute a radial distribution function $g(r)$. Given a test particle is at the origin, $g(r)$ gives the relative probability density of finding another particle at a distance $r$ away, such that $g(r) = 1$ in a non-interacting system. It relates to the radial distribution density $\rho(r)$ as

$$\rho(r) = 4\pi r^2 g(r)/V \qquad (10)$$

with $V$ being the volume of the simulation box, and $4\pi r^2 dr/V$ gives the probability of finding a different particle in a spherical shell of a radius $r$ and thickness $dr$ around the test particle, if the particles were non-interacting. $\rho(r)$ is normalized as $\int_0^{\sqrt{3}L/2} \rho(r)dr = 1$, where $L = \sqrt[3]{V}$ is the side length of the box and $\sqrt{3}L/2$ is the maximal distance between two particle under the periodic boundary condition and minimal image convention[9] (note $\rho(r)$ for $r > L/2$ is unphysical and should not be used). Thus, we modify Eq. (1) to

$$g(r^*) = \frac{\int_{r_-}^{r_+} \rho(r)\, dr}{\int_{r_-}^{r_+} (4\pi r^2/V) \exp[\int_{r^*}^{r} (\log g)'(y)dy]\, dr}.$$

To derive the mean force in a canonical ensemble, where $w(\mathbf{r}) \propto \exp[-\beta U(\mathbf{r}^N)]$, with $\beta$ being the reciprocal temperature, we apply Eq. (5) with a vector field $\mathbf{v}_i = \frac{1}{2}\hat{\mathbf{r}}_{12}(\delta_{1i} - \delta_{2i})$, where $i$ is the particle index. Since $\nabla \cdot \mathbf{v} = 2/r_{12}$ with $r_{12} = |\mathbf{r}_{12}|$, we have $(\log \rho)'(r) = \langle \nabla \cdot \mathbf{v} + \beta \mathbf{v} \cdot \mathbf{F} \rangle_r = 2/r + \frac{1}{2}\beta\langle \hat{\mathbf{r}}_{12} \cdot \mathbf{F}_{12} \rangle_r$, where $\hat{\mathbf{r}}_{12} = \mathbf{r}_{12}/r_{12}$ is the unit displacement vector from particles 2 to 1, $\mathbf{F}_{12} = \mathbf{F}_1 - \mathbf{F}_2$ is the difference between the forces exerted on particles 1 and 2, and the average is evaluated at $r_{12} = r$. By Eq. (10),

$$(\log g)'(r) = (\log \rho)'(r) - 2/r = \tfrac{1}{2}\beta\langle \hat{\mathbf{r}}_{12} \cdot \mathbf{F}_{12} \rangle_r.$$

We simulated the 256 Lennard-Jones system with the density $\rho = 0.7$ under two different temperatures $T_1 = 0.85$ and $T_2 = 0.4$, using the switched potential ($r_s = 2.5$ and $r_c = 3.5$). Velocity rescaling with time step 0.01 was used as the thermostat[7].

After $10^6$ steps of equilibration, we simulated another $10^7$ steps with a time step $dt = 0.002$. We then constructed two samples from this trajectory; for the test sample, we picked 5 frames (from every $2\times 10^6$ steps), and for the reference one, 5000 frames (every 2000 steps). The setting avoided possible sampling inaccuracy to affect the comparison of different methods, as explained in Appendix D. Unless specified otherwise, the test sample was used. Note each frame still contributes $256 \times 255/2 =$ 32640 pairs of particles, although only half ($\pi/6 \approx 0.5$) of them satisfy $r < L/2$. The bin size $\delta r = 0.002$ was used in collecting histogram-like data.

Since this example was also used in the original paper from Adib and Jarzynski[1], it is instructive to compare the two identities. We include the AJ identity [Eq. (20) in reference 1] here for convenience

$$g(r^*) = \frac{e^{-\beta u(r^*)}}{V_{\Omega^*}/V} \left[ \int_{R_1}^{R_{max}} \rho(r) e^{\beta u(r)} dr + \beta \int_0^{R_{max}} \rho(r) \phi(r,r^*) F_r(r) dr \right], \quad (11a)$$

$$\phi(r,r^*) = \frac{e^{\beta u(r)}}{3r^2} \left[ (r^3 - R_1^3)\theta(r - R_1) + (R_1^3 - R_{max}^3)\theta(r - r^*) \right], \quad (11b)$$

where $r$ is the distance from the test particle, $u(r)$ is the pair potential (see Appendix C), $V_{\Omega^*} = (4\pi/3)(R_{max}^3 - R_1^3)$, $R_1$ is a distance close to the repulsion core (we used $R_1 = 1.0$ here), $R_{max}$ is half of the side length of the simulation box, $\theta(x)$ is the step function: 1 if $x > 0$, 0 otherwise, $F_r(r) = \langle \hat{\mathbf{r}} \cdot \mathbf{F}(\mathbf{r}) \rangle_{|\mathbf{r}|=r} + u'(r)$ is the radial mean force of a particle at a distance $r$ away from the test particle, excluding the contribution from the test particle.

As shown in Figure **4**, the fractional identity produced smooth distributions with good agreement with the respective references in both temperatures, despite a relatively small sample size. On the other hand, although the AJ identity also produced smooth distributions, there was an appreciable deviation at the lower temperature $T_2 = 0.4$ from the reference, especially around the principle peak $r \approx 1.1$. The deviation was again not systematic, as it became negligible when the calculation was performed on the reference sample, see Fig. **7**(c) and (d). We note the large deviations from the AJ identity were similar to those observed in the previous example of the volume distribution when the mean force integration was used. They were likely due to an overly large window, as the entire range of $r$, from 0 to $R_{max}$, was used as the window by the AJ identity Eq. (11a). The error was larger at a lower temperature because the mean force changed more drastically there (hence larger mean force fluctuation). By contrast, in the fractional identity case, smaller windows, $\Delta r = 0.14$ for $T_1 = 0.85$ and $\Delta r = 0.09$ for $T_2 = 0.4$ were used according to Eq. (6) ($\gamma = 1.5$), and thus the output was more robust. We also note that the optimal window size shrunk at the lower temperature. The trend is general, since $f_x$ has a component $\beta \mathbf{v} \cdot \mathbf{F}$ in the canonical ensemble, as one increases $\beta$ (or lowers the temperature), the fluctuation grows, and thus the window size shrinks. The example again illustrates the critical influence from the window size.

**III.D Amino acid backbone dihedral angles**

Finally, we compute a joint distribution $\rho(\varphi, \psi)$ of the two backbone dihedrals $\varphi$ and $\psi$ of a glycine dipeptide. Here $\varphi$ and $\psi$ are the C′-N-C$_\alpha$-C and N-C$_\alpha$-C-N′ dihedral angles respectively. We first generalize Eq. (1) to the two dimensional case

$$\rho(\varphi^*, \psi^*) = \frac{\int_{\psi_-}^{\psi_+} \int_{\phi_-}^{\phi_+} \rho(\varphi, \psi) d\phi d\psi}{\int_{\psi_-}^{\psi_+} \int_{\phi_-}^{\phi_+} \exp[\log \rho(\varphi, \psi) - \log \rho(\varphi^*, \psi^*)] d\phi d\psi}.$$

On the denominator, $\log \rho(\varphi, \psi)$ can still be computed from the two mean force components $\frac{\partial}{\partial \varphi} \log \rho(\varphi, \psi)$ and $\frac{\partial}{\partial \psi} \log \rho(\varphi, \psi)$, but not via a direct integration, as the calculation is an overdetermined one (i.e. one distribution but two derivatives). Instead, we constructed $\log \rho(\varphi, \psi)$ in such a way that its two partial derivatives matched the observed values with a minimal overall deviation, see Appendix E.

The mean force are computed as averages as $\frac{\partial}{\partial \varphi} \log \rho(\varphi, \psi) = \langle \beta \mathbf{v}_\varphi \cdot \mathbf{F} \rangle_{\varphi, \psi}$ and $\frac{\partial}{\partial \psi} \log \rho(\varphi, \psi) = \langle \beta \mathbf{v}_\psi \cdot \mathbf{F} \rangle_{\varphi, \psi}$, in which $\beta$ is the temperature, $\mathbf{F}$ is the force, and the two vector fields $\mathbf{v}_\varphi = \nabla_1 \varphi / (\nabla_1 \varphi \cdot \nabla_1 \varphi)$, $\mathbf{v}_\psi = \nabla_4 \psi / (\nabla_4 \psi \cdot \nabla_4 \psi)$. Note by $\nabla_1 \varphi$ and $\nabla_4 \psi$, we mean the gradient components from the atom C′ for $\varphi$, and those from the atom N′ for $\psi$, respectively. The above mean force formulas were free from both the cross correlation and the two divergences $\nabla \cdot \mathbf{u}$ and $\nabla \cdot \mathbf{v}$, see Appendix F for details.

We dissolved the glycine dipeptide in a 32×32×32 Å³ TIP3P[11] water box, and ran the simulation for 36 ns with a time step 1 femtosecond. All chemical bonds of the peptide were allowed to vibrate. A double precision GROMACS 4.5[12] was used as the simulating engine, The velocity rescaling method[7], SETTLE[13], and particle meshed Ewald (PME) sum[14] were used for thermostat, constraints in water molecules and long range electrostatic interaction, respectively. Non-bonded interactions were cutoff at 7 Å and shifted to zero until 8Å. The PME grid spacing was 12Å. Dihedral data were collected every step using a 1°×1° bin.

Due to a relatively large mean force fluctuation, we were only able to use small 4°×4° windows for the fractional identity, according to Eq. (6) (however, in case the window was empty, we expanded the window symmetrically until it included at least one data point). In Fig. **5**, we show that the distribution $\rho(\varphi, \psi)$ from the fractional identity improved the smoothness over that from the normalized histogram. Particularly, the barrier regions, e.g. $\varphi \approx 0°$, $\psi \approx \pm 100°$, were enhanced. Additionally the forbidden band at $\varphi \approx \pm 180°$, where the histogram was simply missing, was now filled by small but finite numbers. On the other hand, peaks at the helical and extended conformations were well preserved. However the overall gain in this case was very modest due to the large mean force fluctuation.

## IV. Conclusions and discussions

In conclusion, we presented an identity, Eqs. (1) and (5), for estimating a general statistical distribution from data collected in molecular simulations. The new identity has broad applications (e.g., to any variable $x$ and any ensemble, easily extended to higher dimensional distributions, etc.), and at the same time offers a robust and precise output.

The general expression for the conjugate force $f_x$ Eq. (5) is also simpler than the conventional $-\beta \, \partial U/\partial x = -\beta \nabla U \cdot \partial \mathbf{r}^N / \partial x$ [15] in that it avoids an inconvenient coordinate transformation in computing $\partial \mathbf{r}^N / \partial x$ by replacing it with a simpler dot product $\beta \mathbf{v} \cdot \mathbf{F}$ plus a divergence $\nabla \cdot \mathbf{v}$. Thus, it is straightforwardly applicable, at least in principle, to an arbitrary $x = X(\mathbf{r}^N)$. Though the computation of the divergence adds computational complexity, it can be sometimes simplified or avoided with a careful choice of the vector field $\mathbf{v}$ (as illustrated in the dihedral example).

We also showed that the window size should be carefully chosen to maximize the benefit from the identity. An overly wide window risks a large error from the mean-force integration, although it usually yields a smoother distribution.

Finally, the method differs from an explicit smoothing method[16] for we do not explicitly assume that the distribution density $\rho(x)$ is smooth here. Although the use of the mean force $(\log \rho)'$ implies a differentiable $\rho(x)$, the mean force can be as oscillatory as an apparent noise. It is however possible that under some approximations the method can be modified for applications where the mean force is unavailable.

## Acknowledgements

We thank Dr. Michael Deem for introducing us to the work of Adib and Jarzynski and for many encouraging and helpful discussions. We also thank Drs. Thomas Woolf, John Straub, Michael Shirts and Thomas Truskett for helpful comments and discussions as well as the referee for critical comments. JM acknowledges support from National Institutes of Health (R01-GM067801), National Science Foundation (MCB-0818353), The Welch Foundation (Q-1512), the Welch Chemistry and Biology Collaborative Grant from John S. Dunn Gulf Coast Consortium for Chemical Genomics and the Faculty Initiatives Fund from Rice University. This work was also supported by the Shared University Grid at Rice funded by NSF under Grant EIA-0216467.

## Appendix A. Alternative derivation of the fractional identity

We first rephrase the work of Adib and Jarzynski in our context. The presentation is given with respect to a general distribution, and thus formally differs from the original one[1], which focused on a radial distribution. Nevertheless, the essences are similar.

First, any function $\rho(x)$ at $x = x^*$ can be evaluated as an integral over $(x_-, x^*)$ as

$$\rho(x^*) = \rho(x^*)\,\phi(x^*) - \rho(x_-)\,\phi(x_-) = \int_{x_-}^{x^*} [\rho(x)\,\phi'(x) + \rho'(x)\,\phi(x)]\,dx,$$

where $\phi(x)$ is an arbitrary function under two conditions, $\phi(x_-) = 0$ and $\phi(x^*) = 1$; we have also converted the difference of $\rho(x)\phi(x)$ to an integral within the range.

The domain of integration can be enlarged to a window $(x_-, x_+)$ that encloses $x^*$ by applying the above equation to two windows $(x_-, x^*)$ and $(x^*, x_+)$ $(x_- < x^* < x_+)$ with different $\phi(x)$'s, and then linearly combining them, see Fig. **6**,

$$\rho(x^*) = \int_{x_-}^{x_+} \rho(x)\,\phi'(x)\,dx + \int_{x_-}^{x_+} \rho'(x)\,\phi(x)\,dx, \tag{A1}$$

where the combined $\phi(x)$ satisfies

$$\begin{cases} \phi(x_+) = \phi(x_-) = 0, \\ \phi(x^{*-}) - \phi(x^{*+}) = 1, \end{cases} \tag{A2}$$

with $\phi(x^{*-})$ and $\phi(x^{*+})$, the values of $\phi(x)$ immediately at the left and right of $x^*$ respectively, serving as coefficients of combination. The function $\phi(x)$ is equivalent to the vector field $\mathbf{u}(\mathbf{r})$ in the original paper[1].

The problem of Eq. (A1) is that it does not always yield a positive output since a negative value of $\rho'(x)$ in the correction $\int_{x_-}^{x_+} \rho'(x)\phi(x)\,dx$ from $\rho'(x)$ can accidentally overthrow the histogram contribution $\int_{x_-}^{x_+} \rho(x)\phi'(x)\,dx$.

We now derive the fractional identity Eq. (1) from Eq. (A1). We start from a simple observation: for any differentiable function $f(x)$ satisfying $f(x^*)=1$, Eq. (A1) applies not only to $\rho(x)$ itself, but also to the product $\rho(x)f(x)$, i.e.,

$$\rho(x^*) = \int_{x_-}^{x_+} [\rho(x)f(x)]\phi'(x)\,dx + \int_{x_-}^{x_+} [\rho(x)f(x)]'\phi(x)\,dx.$$

An arbitrary $f(x)$, or equivalently a reference distribution[1,15], does not guarantee a non-negative output. However, if we choose $f(x)$ such that $\rho(x)f(x)$ a constant, the second term on the right hand side of the above equation vanishes, and

$$\rho(x^*) = \int_{x_-}^{x_+} [\rho(x)f(x)]\phi'(x)\,dx. \qquad (A3)$$

Thus $\rho(x^*)$ is nonnegative as long as $\phi'(x)$ is so. We obtain $f(x)$ from integrating the distribution mean force from the boundary $f(x^*)=1$ as

$$f(x) = \exp\{-[\log\rho(x) - \log\rho(x^*)]\} = \exp\left(-\int_{x^*}^{x} (\log\rho)'(y)\,dy\right).$$

It is easily verified that $\frac{d}{dx}\log f(x) = -\frac{d}{dx}\log\rho(x)$ and hence $\frac{d}{dx}[\rho(x)f(x)] = 0$.

Finally, we determine $\phi(x)$ based on the observation that $\phi'(x)$ acts as a weight in Eq. (A3). To minimize statistical error, $\phi'(x)$ should be inversely proportional to the variance of $\rho(x)f(x)$[9]. For a small window, we assume that the error of $\rho(x)f(x)$ comes mainly from the number of visits $\rho(x)$ instead of $f(x)$. If we assume that the variance of $\rho(x)$ is proportional to $\rho(x)$, i.e., a Poisson distribution[9], then

$\mathrm{Var}[\rho(x)f(x)] = \mathrm{Var}[\rho(x)] f^2(x) \propto \rho(x) f^2(x) \propto f(x)$. $\phi'(x)$ can now be written as $C/f(x)$, where $C$ is a constant determined from Eqs. (A2) as

$$1 = \phi(x^{*-}) - \phi(x_-) + \phi(x_+) - \phi(x^{*+}) = \int_{x_-}^{x_+} \phi'(x)\,dx \text{ (excluding the singularity at } x = x^*).$$

Solving the equation gives $C = 1 \Big/ \int_{x_-}^{x_+} \exp\left[\int_{x^*}^{x} (\log \rho)'(y)\,dy\right] dx$, and Eq. (1) is recovered.

## Appendix B. Improving the mean force

We show how to obtain a precise mean force $(\log \rho)'(x)$. If the second-order derivative $(\log \rho)''(x)$ is available, one can apply an Adib-Jarzynski-like identity (see Appendix A) as:

$$(\log \rho)'(x_0) = \int_{x_-}^{x_+} \left[\varphi'(y)(\log \rho)'(y) + \varphi(y)(\log \rho)''(y)\right] dy, \tag{B1}$$

where $\varphi(x)$ satisfies $\varphi(x_-) = \varphi(x_+) = 0$, $\varphi(x_0 - \delta) - \varphi(x_0 + \delta) = 1$. For simplicity, we particularly use a linear function $\varphi(x) = (x - x_b)/(x_+ - x_-)$, with $x_b = x_-$ if $x < x_0$ or $x_b = x_+$ otherwise. Note the window $(x_-, x_+)$ can be different from that in Eq. (1). By taking the derivative of Eq. (5), we find

$$(\log \rho)''(x) = \left\langle (\Delta f_x)^2 \right\rangle_x + \left\langle \mathbf{v}_x \cdot \nabla f_x \right\rangle_x, \tag{B2}$$

where $\mathbf{v}_x$ denotes the vector field given by Eq. (4) or (4′) for the quantity $x$.

Eqs. (B1) and (B2) are particularly useful in computing the volume distribution, in which a smooth multiplicative correction term $\beta \langle p_c(V) \rangle_V$ can be obtained from the mean force $(\log \rho)'(V) = \beta(\langle p_c(V) \rangle_V - p)$ using the above method.

We list the formulas of $\mathbf{v}_x \cdot \nabla f_x$ for the first three examples in the Examples.

For the potential energy distribution in the canonical ensemble, $f_U = \nabla \cdot \mathbf{v} - \beta$, and

$$\mathbf{v}_U \cdot \nabla f_U = -\frac{\mathbf{F} \cdot \nabla\nabla^2 U}{(\mathbf{F} \cdot \mathbf{F})^2} - \frac{\nabla^2 U\, M + \mathbf{h} \cdot \mathbf{h} - 2\mathbf{FFF} \vdots \nabla\nabla\nabla U}{(\mathbf{F} \cdot \mathbf{F})^3} + \frac{2M^2}{(\mathbf{F} \cdot \mathbf{F})^4},$$

where $\mathbf{F} = -\nabla U$ is the force, $M \equiv 2\mathbf{F} \cdot \nabla\nabla U \cdot \mathbf{F}$, $\mathbf{h} = 2\nabla\nabla U \cdot \mathbf{F}$, and the "$\vdots$" denotes the triple dot product between two tensors of order three. For the microcanonical ensemble, we add an additional term $-(\tfrac{1}{2}N_f - 1)/(E_{\text{tot}} - U)^2$ to the above formula.

For the volume distribution, $\mathbf{v} \cdot \nabla \to \partial/\partial V$, $f_V = N/V - \beta\, \partial U/\partial V$, and

$$\begin{aligned}
\mathbf{v}_V \cdot \nabla f_V &= \frac{\partial f_V}{\partial V} = -\frac{N}{V^2} - \frac{\beta\, \partial^2 U}{\partial V^2} \\
&= -\frac{N}{V^2} - \frac{\beta \sum_{(m,n)} [\psi(r_{mn}) r_{mn}^4 - \phi(r_{mn}) r_{mn}^2]}{9V^2},
\end{aligned}$$

where on the second line, we have assumed the potential energy $U$ is a sum over particle pairs $(m,n)$ for the molecular potential $u(r)$, i.e., $U = \sum_{(m,n)} u(r_{mn})$, and $\phi(r) = u'(r)/r$, $\psi(r) = \phi'(r)/r$.

For the radial distribution function of a pair of particles 1 and 2,

$$\begin{aligned}
\mathbf{v}_r \cdot \nabla f_r = &-\beta[\psi(r_{12}) r_{12}^2 + \phi(r_{12})] \\
&- \beta \sum_{k \neq 1,2} \left[ \frac{\psi(r_{1k})(\mathbf{r}_{1k} \cdot \mathbf{r}_{12})^2 + \psi(r_{2k})(\mathbf{r}_{2k} \cdot \mathbf{r}_{12})^2}{4r_{12}^2} + \frac{\phi(r_{1k}) + \phi(r_{2k})}{4} \right],
\end{aligned}$$

where $\mathbf{r}_{jk}$ is the displacement vector from $k$ to $j$, and $r_{jk} = |\mathbf{r}_{jk}|$.

## Appendix C. Switched Lennard-Jones potential

The Lennard-Jones potential $u(r) = 4\varepsilon[(\sigma/r)^{12} - (\sigma/r)^6]$ is switched at $r = r_s$ to a polynomial $\sum_{k=4}^{7} a_k (r - r_c)^k$ and extended to zero at $r = r_c$. For simplicity, we assume the reduced unit, whereby both the energy unit $\varepsilon$ and diameter $\sigma$ are 1.0. In our example of the energy distribution, we needed the potential and the first three derivatives to be continuous, since the derivative of the dynamic temperature requires up to the third-order derivative of the potential. The continuity at $r = r_c$ is guaranteed by the first four vanishing coefficients. To ensure the continuity up to third-order derivatives at $r = r_s$, the following parameters are used

$$a_4 = 4\,(35\,r_s^3 A + 90\,r_s^2 \Delta r\, B + 15\,r_s\, \Delta r^2 C + 28\,\Delta r^3 D)/(\Delta r^4\, r_s^{15}),$$
$$a_5 = -24\,(14\,r_s^3 A + 39\,r_s^2 \Delta r\, B + 7\,r_s\, \Delta r^2 C + 14\,\Delta r^3 D)/(\Delta r^5\, r_s^{15}),$$
$$a_6 = 4\,(70\,r_s^3 A + 204\,r_s^2 \Delta r\, B + 39\,r_s\, \Delta r^2 C + 84\,\Delta r^3 D)/(\Delta r^6\, r_s^{15}),$$
$$a_7 = -16\,(5\,r_s^3 A + 15\,r_s^2 \Delta r\, B + 3\,r_s\, \Delta r^2 C + 7\,\Delta r^3 D)/(\Delta r^7\, r_s^{15}),$$

where, $\Delta r = r_c - r_s$, $A = 1 - r_s^6$, $B = 2 - r_s^6$, $C = 26 - 7r_s^6$, $D = 13 - 2r_s^6$.

## Appendix D. Reference distributions

In each of Sections III.A, III.B and III.C, we prepared both the test and reference samples from the same trajectory but with different sampling rates of picking frames. Since frames in the test sample were just a subset of those in the reference one, any sampling inaccuracy, e.g., due to insufficient equilibration or sampling, would be shared by both samples, and thus would not affect the comparison of different methods. We also emphasize that, in either sample, the numbers of sampling points available to different methods were exactly the same.

We have used the fractional identity to produce the reference distributions, in Figs. **2**, **3**, and **4**, which might be unfair to the alternative identities. However, due to the large size of the reference sample, the reference distributions were insensitive to what method we choose. In Fig. **7**, we show that the results from the fractional identity agreed well with the properly normalized histograms, as well as those from alternative identities [compare Figs. **2** with **7**(a), **3** with **7**(b), and **4** with **7**(c)-(d)]. The latter comparison also shows the alternative identities are unbiased, although less stable in handling the smaller samples. Note we used the WHAM version in producing the reference distribution in Fig. **3**(a), which further improved over the one in Fig. **7**(a) at the edges.

## Appendix E. Potential from the two-dimensional mean force

We determine a two-dimensional "potential" $u(x, y) \equiv \log \rho$ from the two mean force components $f = \partial u/\partial x$ and $g = \partial u/\partial y$. We only seek the "best fit" solution below, for the problem is overdetermined: the number of unknowns $u$ is only half of that of the equations ($f$ and $g$).

We assume the potential is set up on a two-dimensional grid of $N \times M$ cells of size $\delta x \times \delta y$. For the cell at $(n, m)$, where $n = 1, 2, \ldots N$, $m = 1, 2, \ldots M$, we wish the mean force value $f_{n,m}$ to match the value from discretely differentiating the $u$ values at four cell corners $(u_{n+1,m} + u_{n+1,m+1} - u_{n,m} - u_{n,m+1})/(2\delta x)$ (and similarly for $g_{n,m}$). Therefore we minimize the following action $S$,

$$S = \frac{1}{2}\sum_{n,m}\left[\left(\frac{u_{n+1,m}+u_{n+1,m+1}-u_{n,m}-u_{n,m+1}}{2}-f_{n,m}\,\delta x\right)^2+\left(\frac{u_{n,m+1}+u_{n+1,m+1}-u_{n,m}-u_{n+1,m}}{2}-g_{n,m}\,\delta y\right)^2\right].$$

At the minimum, we have $\partial S/\partial u_{n,m}=0$ for every $u_{n,m}$, i.e.,

$$u_{n,m}-\frac{u_{n-1,m-1}+u_{n-1,m+1}+u_{n+1,m-1}+u_{n+1,m+1}}{4}$$
$$=\frac{f_{n-1,m-1}+f_{n-1,m}-f_{n,m-1}-f_{n,m}}{4}\delta x+\frac{g_{n-1,m-1}+g_{n,m-1}-g_{n-1,m}-g_{n,m}}{4}\delta y$$

The set of linear equations can be solved using Fourier transform. With

$$\tilde{u}_{k,l}=\sum_{n,m}u_{n,m}\exp[-2\pi i(kn/N+lm/M)]\ (\tilde{f}_{k,l}\text{ and }\tilde{g}_{k,l}\text{ similarly defined}),\text{ we have}$$

$$\tilde{u}_{k,l}=\frac{-i\exp\left[-\pi i\left(\frac{k}{N}+\frac{l}{M}\right)\right]\left[\sin\left(\frac{\pi k}{N}\right)\cos\left(\frac{\pi l}{M}\right)\tilde{f}_{k,l}\,\delta x+\cos\left(\frac{\pi k}{N}\right)\sin\left(\frac{\pi l}{M}\right)\tilde{g}_{k,l}\,\delta y\right]}{1-\cos\left(\frac{2\pi k}{N}\right)\cos\left(\frac{2\pi l}{M}\right)}.$$

A final inverse Fourier transform yields the desired potential $u_{n,m}$ in the real space.

## Appendix F. Joint dihedral distribution

Here we derive a general mean force formula [analogous to Eq. (5)] for the joint dihedral distribution. We start from the definition

$$\rho(\varphi,\psi)=\int\delta(\varphi-\Phi(\mathbf{r}^N))\delta(\psi-\Psi(\mathbf{r}^N))w(\mathbf{r}^N)d\mathbf{r}^N.$$

Following a similar derivation leading to Eq. (5), we have

$$\frac{\partial}{\partial\varphi}\rho(\varphi,\psi)=\int-\mathbf{v}_\varphi\cdot\nabla[\delta(\varphi-\Phi(\mathbf{r}^N))]\delta(\psi-\Psi(\mathbf{r}^N))w(\mathbf{r}^N)d\mathbf{r}^N$$
$$=\int\delta(\varphi-\Phi(\mathbf{r}^N))\nabla\cdot[\mathbf{v}_\varphi\,\delta(\psi-\Psi(\mathbf{r}^N))w(\mathbf{r}^N)]d\mathbf{r}^N$$
$$=\int(\nabla\cdot\mathbf{v}_\varphi+\mathbf{v}_\varphi\cdot\nabla\log w)\delta(\varphi-\Phi)\delta(\psi-\Psi)w\,d\mathbf{r}^N$$
$$-\frac{\partial}{\partial\psi}\int(\mathbf{v}_\varphi\cdot\nabla\Psi)\delta(\varphi-\Phi)\delta(\psi-\Psi)w\,d\mathbf{r}^N.$$

where $\mathbf{v}_\varphi$ is a vector field satisfying $\mathbf{v}_\varphi \cdot \nabla\Phi = 1$. We note that the last integral, which is hard to evaluate, is avoidable if $\mathbf{v}_\varphi \cdot \nabla\Psi = 0$. Thus we shall generalize Eq. (4′) to satisfy both $\mathbf{v}_\varphi \cdot \nabla\Phi = 1$ and $\mathbf{v}_\varphi \cdot \nabla\Psi = 0$ as

$$\mathbf{v}_\varphi = \frac{(\nabla\Psi \cdot \mathbf{Y}_\psi)\mathbf{Y}_\varphi - (\nabla\Psi \cdot \mathbf{Y}_\varphi)\mathbf{Y}_\psi}{(\nabla\Psi \cdot \mathbf{Y}_\psi)(\nabla\Phi \cdot \mathbf{Y}_\varphi) - (\nabla\Psi \cdot \mathbf{Y}_\varphi)(\nabla\Phi \cdot \mathbf{Y}_\psi)}, \tag{F1a}$$

where $\mathbf{Y}_\varphi$ and $\mathbf{Y}_\psi$ are such vector fields that $\nabla\Phi \cdot \mathbf{Y}_\varphi \neq 0$ and $\nabla\Psi \cdot \mathbf{Y}_\psi \neq 0$. Similarly

$$\mathbf{v}_\psi = \frac{(\nabla\Phi \cdot \mathbf{Y}_\varphi)\mathbf{Y}_\psi - (\nabla\Phi \cdot \mathbf{Y}_\psi)\mathbf{Y}_\varphi}{(\nabla\Phi \cdot \mathbf{Y}_\varphi)(\nabla\Psi \cdot \mathbf{Y}_\psi) - (\nabla\Phi \cdot \mathbf{Y}_\psi)(\nabla\Psi \cdot \mathbf{Y}_\varphi)}. \tag{F1b}$$

If the two dihedrals are decoupled $\nabla\Phi \cdot \mathbf{Y}_\psi = \nabla\Psi \cdot \mathbf{Y}_\varphi = 0$, Eqs. (F1a) and (F1b) are then reduced to Eq. (4′), i.e., $\mathbf{v}_\varphi = \mathbf{Y}_\varphi / (\nabla\Phi \cdot \mathbf{Y}_\varphi)$ and $\mathbf{v}_\psi = \mathbf{Y}_\Psi / (\nabla\Psi \cdot \mathbf{Y}_\Psi)$. This condition is satisfied if $\mathbf{Y}_\varphi = \nabla_1\Phi$ and $\mathbf{Y}_\psi = \nabla_4\Psi$, i.e., the components from the atom C′ in C′-N-C$_\alpha$-C for $\nabla\Phi$, and those from the atom N′ in N-C$_\alpha$-C-N′ for $\nabla\Psi$.

We can also show that in this case $\nabla \cdot \mathbf{v}_\varphi = \nabla \cdot \mathbf{v}_\psi = 0$ (i.e., if only the 1, 4 atoms are involved in $\mathbf{Y}_\varphi$ and $\mathbf{Y}_\psi$). Using $\Phi$ as example, we first label the C′, N, C$_\alpha$ and C atoms by 1, 2, 3, 4 respectively, and express the cosine of the dihedral as $\cos\Phi = \hat{\mathbf{m}} \cdot \hat{\mathbf{n}}$, where $\hat{\mathbf{m}} = \mathbf{m}/m$ and $\hat{\mathbf{n}} = \mathbf{n}/n$ are the unit vectors of the planes 1-2-3 and 2-3-4 respectively, with $\mathbf{m} = \mathbf{r}_{12} \times \mathbf{r}_{32}$, $m = |\mathbf{m}|$, $\mathbf{n} = \mathbf{r}_{32} \times \mathbf{r}_{34}$ and $n = |\mathbf{n}|$. The components of the gradient are listed below,

$$\begin{aligned}\nabla_1\Phi &= r_{32}\,\hat{\mathbf{m}}/m, \\ \nabla_2\Phi &= (\mathbf{r}_{13} \cdot \hat{\mathbf{r}}_{32})\hat{\mathbf{m}}/m - (\mathbf{r}_{43} \cdot \hat{\mathbf{r}}_{32})\hat{\mathbf{n}}/n, \\ \nabla_3\Phi &= (\mathbf{r}_{42} \cdot \hat{\mathbf{r}}_{32})\hat{\mathbf{n}}/n - (\mathbf{r}_{12} \cdot \hat{\mathbf{r}}_{32})\hat{\mathbf{m}}/m, \\ \nabla_4\Phi &= -r_{32}\,\hat{\mathbf{n}}/n.\end{aligned} \tag{F2}$$

For completeness, we sketch a geometric derivation of Eqs. (F2). First, the gradient of particle 1 must be parallel to $\hat{\mathbf{m}}$, since a displacement within the plane 1-2-3 leaves the plane vector $\hat{\mathbf{m}}$ and hence $\cos\Phi$ unchanged. The magnitude of the gradient is the inverse of the perpendicular distance from particle 1 to the line connecting particles 2 and 3, or $|\nabla_1\Phi| = r_{32}/m$. Thus $\nabla_1\Phi = (r_{32}/m)\hat{\mathbf{m}}$. Similarly, $\nabla_4\Phi = -r_{32}\hat{\mathbf{n}}/n$. The equation for $\nabla_2\Phi$ can be derived from that the dihedral is invariant upon a rotation around any axis $\mathbf{h}$ passing through particle 3, i.e., $\sum_{a=1,2,4}\nabla_a\Phi\cdot(\mathbf{h}\times\mathbf{r}_{a3}) = 0$. By using $\mathbf{h} = \mathbf{m}$ and $\mathbf{n}$, and solving the equations for $\nabla_2\Phi$, we reach the second equation. The equation for $\nabla_3\Phi$ follows from the invariance of the dihedral upon any translation $\mathbf{d}$,

$\sum_{a=1,2,3,4}\nabla_a\Phi\cdot\mathbf{d} = 0$, or $\nabla_3\Phi = -(\nabla_1\Phi + \nabla_2\Phi + \nabla_4\Phi)$.

Now if $\mathbf{Y}_\varphi = \nabla_1\Phi$, then $\mathbf{v}_{\varphi,i} = (m/r_{32})\hat{\mathbf{m}}\delta_{i1}$. Thus the gradient flow forms concentric circles around the axis connecting particles 2 and 3. From Gauss's law, we must have $\nabla\cdot\mathbf{v}_\varphi = 0$ for the flow has no source or sink. We thus reach the formulas given in the main text for the canonical ensemble: $\frac{\partial}{\partial\varphi}\log\rho(\varphi,\psi) = \langle\beta\mathbf{v}_\varphi\cdot\mathbf{F}\rangle_{\varphi,\psi}$ (and similarly $\frac{\partial}{\partial\psi}\log\rho(\varphi,\psi) = \langle\beta\mathbf{v}_\psi\cdot\mathbf{F}\rangle_{\varphi,\psi}$).

**Figure Captions**

**FIG. 1.** The key of the fractional identity is to represent the ratio of a histogram sum $\int_{x_-}^{x_+} \rho(x)dx$ (shaded area) to the distribution density $\rho(x^*)$ as an integral of the "mean force" $(\log \rho)'(x)$. We then measure the mean force and use its integral to divide the observed histogram sum to obtain an unbiased estimate of $\rho(x^*)$. As both the histogram sum and ratio involves data from a window instead of a single bin, the reduced uncertainty makes the resulting distribution smoother.

**FIG. 2.** The potential energy distribution. **(a)** Comparison of the distributions from the histogram (gray dotted lines), the fractional identity [Eq. (1), blue circles and dashed line]; and the Adib-Jarzynski (AJ) like identity [Eq. (A1), red squares and dot-dashed line]. Data from the same test sample were used for all three; symbols were plotted with a spacing $\Delta U = 5.0$ to avoid cluttering. The reference curve (black line) was computed from the reference sample (same trajectory, higher sampling rate). Lines were plotted with a spacing equal to the bin size $\delta U = 0.1$. Vertical lines at the right edge were due to negative or small output from the AJ identity. **(b)** Error measured from the KS difference as a function of window size $\Delta U$. **(c)** Error measured from the entropic distance. **(d)** The distribution from the weighted histogram analysis method (WHAM, gray dotted line), compared with an improved version using the fractional identity [Eq. (9), blue circles, dashed line]. The styles were similar to those in Panel **(a)**. Vertical dotted lines at the edges were due to empty output from WHAM. **(e)** Comparison of the distributions from a canonical ensemble (blue) and a microcanonical one (red).

**FIG. 3.** The volume distributions (a) $\rho(V)$ and (b) $\hat{\rho}(V)$ (adjusted) at $p = 0.115$ and $T = 1.24$. Gray line in (a): the histogram; blue circles and dashed lines: the fractional identity; red squares and dot-dashed lines: pure mean-force integration (the limiting case of an infinite window). Inset of (a): the window size. Data from the same test sample were used for all three; symbol spacing $\Delta V$ was 50.0 to avoid cluttering. The reference curves (black lines) were computed from the reference sample (same trajectory, higher sampling rate), adjusted in (b). Lines were plotted with a spacing equal to the bin size $\delta V = 1.0$.

**FIG. 4.** The radial distribution function $g(r)$. **(a)** $T_1 = 0.85$; **(b)** $T_2 = 0.4$. Gray line: the histogram; blue circles and dashed line: the fractional identity; red squares and dot-dashed line: the original Adib-Jarzynski (AJ) identity (data from the same test sample were used for all three; symbol spacing $\Delta r$ was 0.1 to avoid cluttering). The reference curve (black line) was computed from the reference sample (same trajectory, higher sampling rate). Lines were plotted with a spacing equal to the bin size $\delta r = 0.002$.

**FIG. 5.** The joint distribution of the two backbone dihedrals in the glycine dipeptide. **(a)** The histogram; **(b)** the fractional identity.

**FIG. 6.** The auxiliary function $\phi(x)$ (solid) and its derivative $\phi'(x)$ (dashed, excluding the $\delta$-function at $x = x^*$) employed by the Adib-Jarzynski-like identity.

**FIG. 7.** Distributions computed from the reference samples: **(a)** the energy distributions; **(b)** the volume distributions; **(c)** and **(d)** the radial distribution functions at $T_1 = 0.85$ and $T_2 = 0.4$ respectively. Gray dotted lines: histograms; blue solid lines: fractional identity; red dashed lines: alternative identities, as those in Figs. **2**, **3**, and **4**.

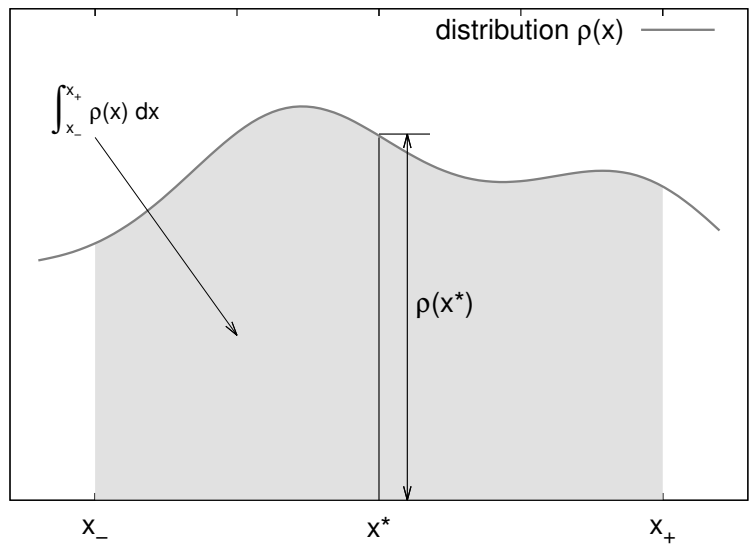

Figure 1:

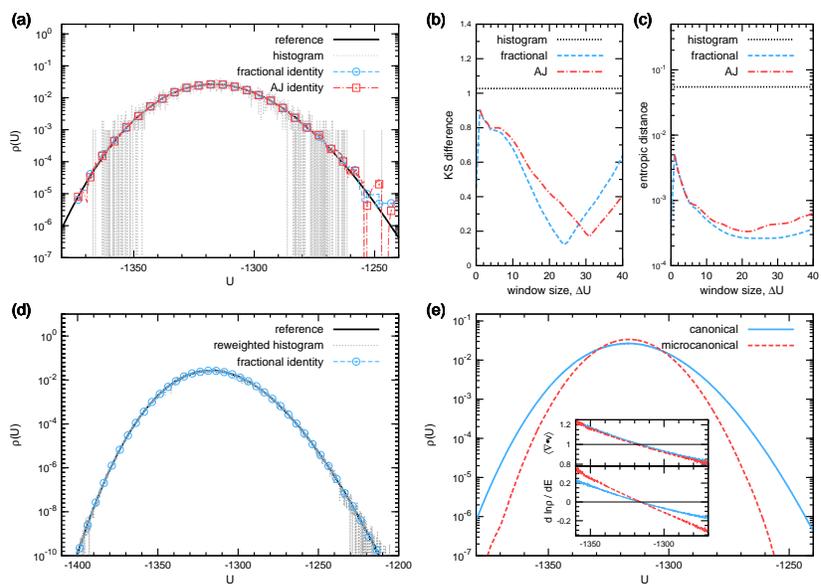

Figure 2:

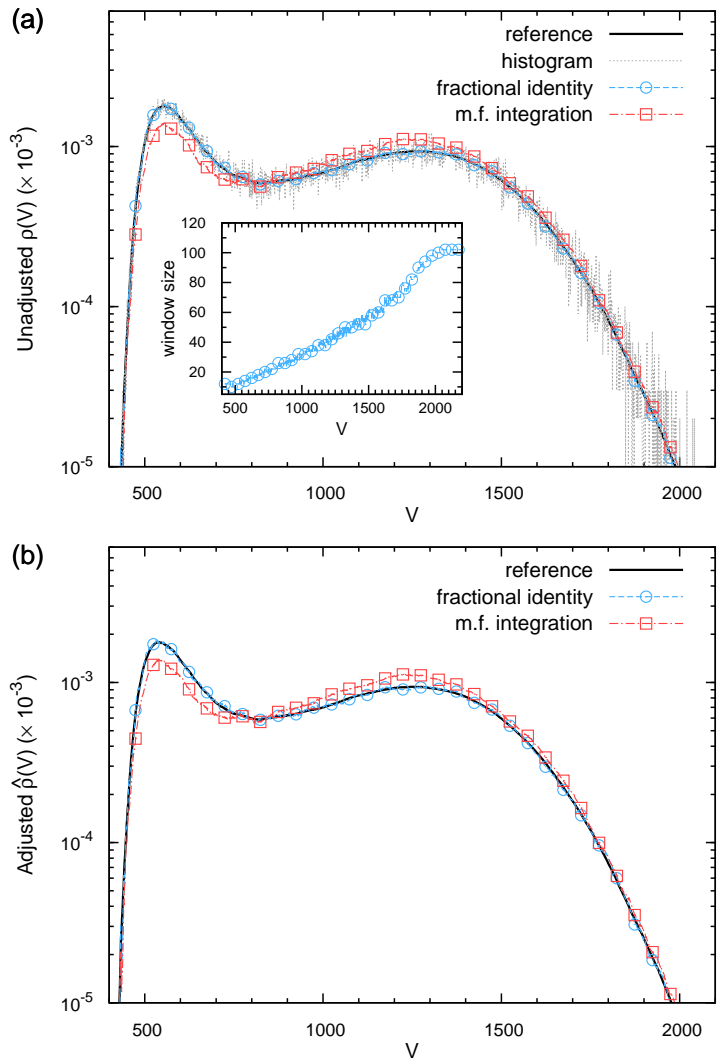

Figure 3:

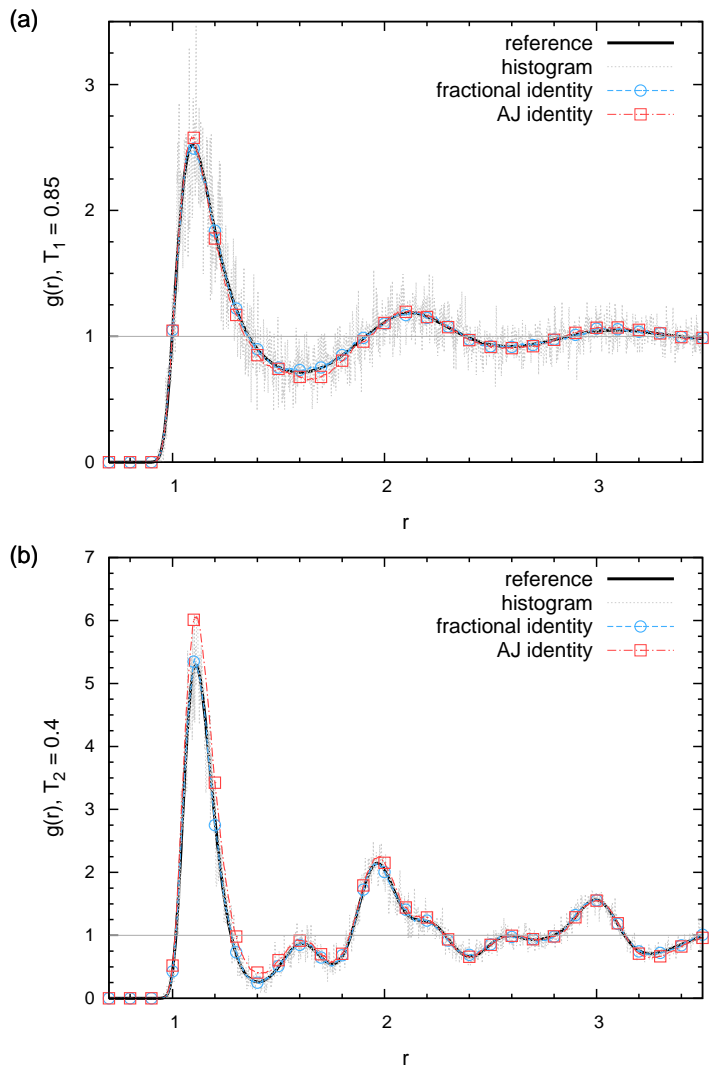

Figure 4:

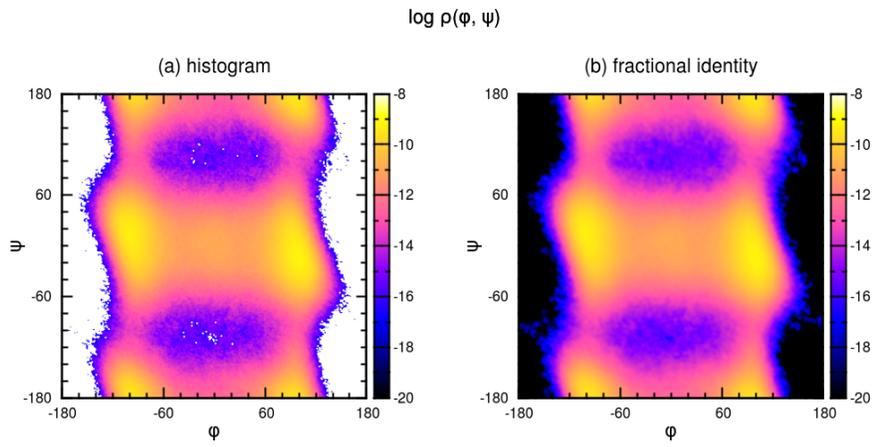

Figure 5:

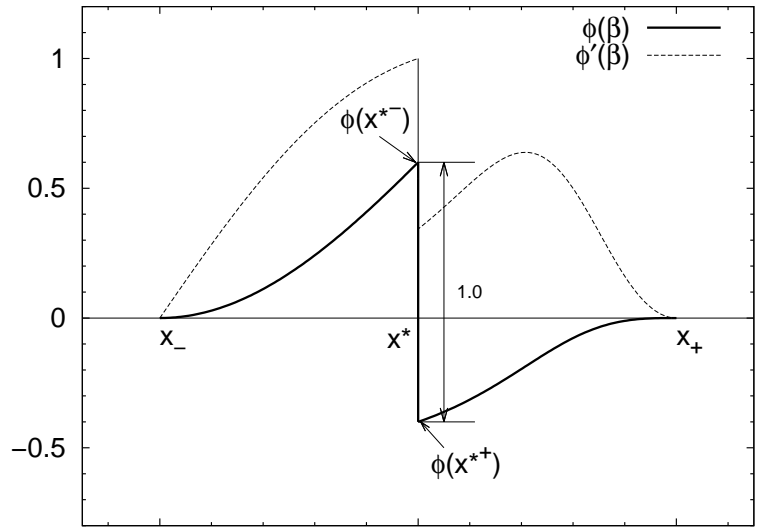

Figure 6:

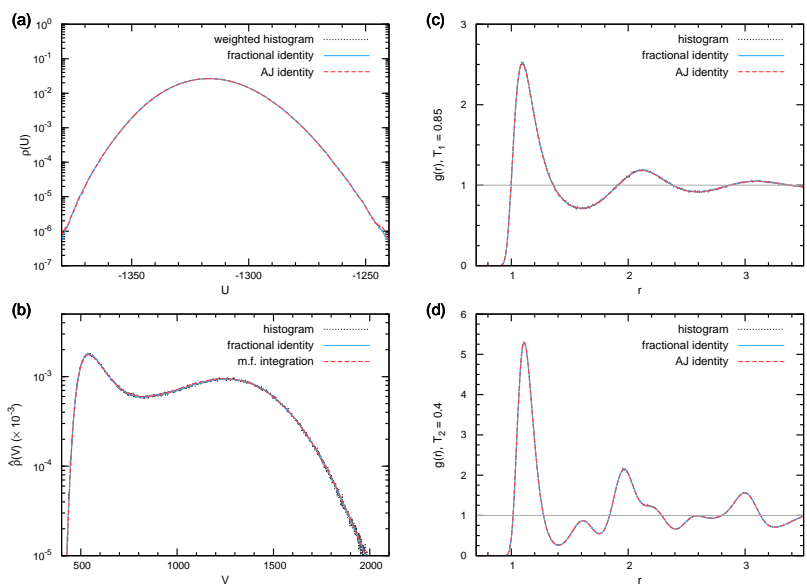

Figure 7: